# Peregrine rogue waves in the nonlocal nonlinear Schrödinger equation with parity-time symmetric self-induced potential

Samit Kumar Gupta [a, *]

[*] samit.kumar.gupta@gmail.com

[a] Department of Physics, Indian Institute of Technology Guwahati, Guwahati 781 039, India

**Abstract -** In this work, based on the recently proposed (Phys. Rev. Lett. **110** (2013) 064105) continuous nonlocal nonlinear Schrödinger system with parity-time symmetric Kerr nonlinearity (PTNLSE), a numerical investigation has been carried out for two first order Peregrine solitons as the initial ansatz. Peregrine soliton, as an exact solution to the PTNLSE, evokes a very potent question: what effects does the interaction of two first order Peregrine solitons have on the overall optical field dynamics. Upon numerical computation, we observe the appearance of Kuznetsov-Ma (KM) soliton trains in the unbroken PT-phase when the initial Peregrine solitons are in phase. In the out of phase condition, it shows repulsive nonlinear waves. Quite interestingly, our study shows that within a specific range of the interval factor in the transverse co-ordinate there exists a string of high intensity well-localized Peregrine rogue waves in the PT unbroken phase. We note that the interval factor as well as the transverse shift parameter play important roles in the nonlinear interaction and evolution dynamics of the optical fields. This could be important in developing fundamental understanding of nonlocal non-Hermitian NLSE systems and dynamic wave localization behaviors.

**1. Introduction**

A wide class of non-Hermitian Hamiltonians obeying parity-time (PT)-symmetry can possess entirely real eigenspectra, a pioneering work put forward by Bender and Boettcher [1]. It followed by an extensive amount of research activities both in theoretical and experimental fronts [2-34]. The Hamiltonian $H = -\frac{1}{2}\frac{d^2}{dx^2} + V(x)$ where $V(x) = V_R(x) + i\, V_I(x)$, is PT-symmetric subject to the condition that the complex potential $V(x)$ satisfies the following relationship: $V^*(-x) = V(x)\, i.e.\, V_R(-x) = V_R(x)$ and $V_I(-x) = -V_I(x)$. The regime with real eigen spectrum is called unbroken PT-phase regime, whereas the spectrum with complex or imaginary eigenvalues is known as broken PT-phase. In the latter case, the imaginary part of the potential $V_I(x)$ goes beyond a critical value of the system to enter into the broken PT-phase. Recently, nonlinear Schrödinger systems with PT-symmetric potentials ($i\psi_z + \frac{1}{2}\psi_{xx} + V(x)\psi + |\psi|^2\psi = 0$) have been studied rigorously with novel effects and possibilities [20-30]. In an NLSE system, an effective linear potential which in general may not be PT-symmetric, is induced by the Kerr nonlinearity. The NLSE system governed by the so-called PT potential can observe the PT instability in its wave dynamics once the imaginary component of the potential goes beyond a certain critical value. In a seminal paper, Ablowitz and Musslimani [31] have proposed an alternative class of fully integrable (since it possesses the Lax pair and an infinite number of conserved quantities) highly nonlocal nonlinear Schrödinger equation. Here the standard third-order nonlinearity $|\psi|^2\psi$ is replaced by its PT-symmetric form $\psi(x,z)\psi^*(-x,z)\psi(x,z)$ and the corresponding NLSE obtained is called PTNLSE. In another work, Christodoulides et al. have considered continuous and discrete Schrödinger systems exhibiting parity-time (PT) symmetric nonlinearity [32], which shows that the system possesses simultaneous bright and dark soliton solutions and that the shift in the transverse co-ordinate '$x$' results in the PT-symmetry breaking. In this connection, there are a number of theoretical investigations carried out in PTNLSE systems including, dark and anti-dark soliton interactions in the nonlocal nonlinear Schrödinger equation with self-induced parity-time-symmetric potential [33], periodic and hyperbolic soliton solutions in nonlocal PT-symmetric equations [34], dynamics of first and higher order rational solitons in PTNLSE model [35, 36], analytic solutions and interaction dynamics in discrete PTNLSE [37], exact solutions and symmetries [38], collisional dynamics of bright and dark solitons [39]. On the other side, Peregrine solitons [40] is a limiting case of a wide range of solutions to the nonlinear Schrödinger equations including the transverse co-ordinate periodic Akhmediev breather [40-42] or the longitudinal co-ordinate periodic Kuznetsov-Ma (KM) breather [41, 43]. Theoretically predicted in 1983 [44] it was not until 2010 that it was experimentally demonstrated in nonlinear fiber optics [43]. Due to its doubly localized behavior, Peregrine solitons received a great deal of attention afterwards. Few among them are: Peregrine solitons in a multi-component plasma with negative ions [45], Peregrine soliton generation and breakup in the standard telecommunications fiber [46], interaction of Peregrine solitons [47], breather-like solitons extracted from the Peregrine rogue wave [48], Peregrine solitons and algebraic soliton pairs in Kerr media [49] and so on. It is interesting to note that Peregrine soliton is taken as a rogue-wave prototype [50] for its closely related rogue-wave dynamics. Now originally studied in hydro-dynamical and other systems [45, 51-53] rogue waves are large amplitude waves finding its extension in optical systems too [54-56].

---

In this work, we have considered an initial excitation condition consisting of two in phase or out of phase Peregrine solitons to the focusing nonlinear Schrödinger equation with anomalous dispersion with parity-time (PT) symmetric nonlinearity. The study aims at numerically investigating the interaction dynamics of these first order Peregrine solitons in the in phase and out of phase conditions in PTNLSE model to look for interesting nonlinear solutions which could be fundamentally important.

The article is arranged as follows. In section 2 the theoretical model has been described. In section 3 the numerical simulation and analysis part of the optical fields dynamics has been elucidated including various nonlinear waves solutions in sub-section 3.1 for in phase and out of phase conditions. The effects of the interval factor on the optical fields dynamics have been discussed in sub-section 3.2 followed by the effects of the transverse shift parameter on the optical field dynamics in sub-section 3.3. The conclusions are drawn along with pertinent discussions in section 4.

## 2. Theoretical model

In this work, we are considering the nonlocal (since the evolution dynamics of the optical field at '$x$' necessitates information at its mirror image '$-x$' [32]) nonlinear Schrodinger equation, in normalized units, where the standard third order nonlinearity $|\psi(x,z)|^2 \psi(x,z)$ is replaced with its PT symmetric counterpart $\psi(x,z)\psi^*(-x,z)\psi(x,z)$ to obtain:

$$i\psi_z + \frac{1}{2}\psi_{xx} + \psi(x,z)\psi^*(-x,z)\psi(x,z) = 0 \quad (1)$$

Here $\psi(x,z)$ is the dimensionless field with $x$ and $z$ being the normalized distance and time. If the total optical power is denoted by $P = \int_{-\infty}^{\infty} dx\, |\psi|^2$, then it satisfies $\frac{dP}{dz} = \int_{-\infty}^{\infty} dx\, |\psi|^2 [\psi\psi^*(-x,z) - \psi^*\psi(-x,z)]$, which essentially signifies that Eq. (1) represents non-Hermitian system [32]. It may be useful to note few of the infinite number of constants of motion, namely the quasi-power $Q = \int_{-\infty}^{\infty} dx\, \psi(x,z)\psi^*(-x,z)$ and the Hamiltonian $H = \int_{-\infty}^{\infty} dx\, [\psi_x(x,z)\psi_x(-x,z) - \psi^2(x,z)\psi^{*2}(-x,z)]$ [32]. It is straightforward to show that Eq. (1) possesses the following solitons on finite background (SFB) solutions:

$$\psi(x,z) = \left[\frac{(1-4a)\cosh(bz) + \sqrt{2a}\cos(\Omega x) + i\, b\, \sinh(bz)}{\sqrt{2a}\cos(\Omega x) - \cosh(bz)}\right] \quad (2)$$

where $\Omega$ is the dimensionless spatial modulation frequency, $a = 1/2(1 - \Omega^2/4)$ with $0 < a < 1/2$ determines the frequencies experiencing gain and $b = \sqrt{\{8a(1-2a)\}}$ is the instability growth parameter. For $a \to 1/2$, the above solution reduces to the rational soliton form, i.e. standard first order Peregrine soliton:

$$\psi(x,z) = \left[1 - \frac{4(1+2iz)}{1+4x^2+4z^2}\right] e^{iz} \quad (3)$$

It is worth noting that the parameterized family of Peregrine solitons scaled by the spectral parameter $\omega$ can be represented by:

$$\psi(x,z) = \sqrt{|\omega|} \left[1 - \frac{4(1+2iz)}{1+4|\omega|x^2+4\omega^2 z^2}\right] e^{-i\omega z} \quad (4)$$

As is obvious, Eq. (4) reduces to Eq. (3) when $\omega = -1$.

Note that the co-ordinates $(x,z)$ has been replaced identically by $(X,Z)$ in the following figures for better visualization.

## 3. Numerical simulations and analysis

In order to analyze the interaction of two Peregrine solitons in the continuous PTNLSE model numerically, we have taken the initial excitation at the incident port at $z = 0$ following the translation invariance in transverse co-ordinate and scale invariance in amplitude as:

$$\psi(x,0) = \left[1 - \frac{4}{1+4(x-\varepsilon_+)^2}\right] + e^{il\pi}\left[1 - \frac{4}{1+4(x+\varepsilon_-)^2}\right] \quad (5)$$

In a recent work [32] it is shown that introduction of the shift $\varepsilon$, in the transverse co-ordinate $x$ may give rise to instability in the wave dynamics. This happens because the self-induced potential $V(x,z)$ starts to have imaginary contribution, due to transverse shift, which once goes beyond a critical value results in instability of the wave dynamics. This motivates us to consider various cases with regard to broken and unbroken PT-phases. $e^{il\pi}$ is the phase factor of the two Peregrine solitons where $l = 0$ and $l = 1$ correspond to the in phase and out of phase conditions respectively. Eqn. (1) has been solved numerically by the split-operator method [57]. Here we define $\varepsilon_\pm = \varepsilon_{loc} \pm \varepsilon_{tsp}$ where $\varepsilon_{loc}$ is the initial location of the two input Peregrine solitons known as the interval factor and $\varepsilon_{tsp}$ is the transverse shift parameter. In this work, except the sub-section 3.3 (where $\varepsilon_{tsp} \neq 0$) we have considered the case when $\varepsilon_+ = \varepsilon_-$ which essentially means $\varepsilon_{tsp} = 0$. For simplicity we keep throughout the paper $\varepsilon_+ = \varepsilon_- = \varepsilon_{loc} = \varepsilon$.

### 3.1. KM soliton trains, Peregrine rogue waves and repelling nonlinear waves

We first consider the case when the two initial Peregrine solitons are in phase as shown in Fig. 1 (a), (b). In the absence of any transverse shift in and in the unbroken PT phase we can see from Fig.1 (a)-(b), the existence of Kuznetsov-Ma (KM) solitons periodic in the evolution variable $z$.

Now we know that when the modulation parameter $a$ (see Eqn. (2)) approaches $a \to 1/2$, it gives rise to the Peregrine soliton, which has been shown to be an exact solution of PTNLSE. In Fig. 2(a)-(b), when the transverse shift parameter $\varepsilon_{tsp}$ is zero, after a certain propagation distance the optical fields dynamics results into dynamic localization. The single Peregrine rogue wave solution is observed. We note that with further increase in $z$ it becomes unstable which is inherent characteristic of Peregrine rogue waves. Such single near-ideal Peregrine rogue wave has been shown to exist with similar type of initial excitation condition in NLSE systems (Eq. (5)) as described in [59]. Fig. 1 (c), (d) elucidate the case of two out of phase first order Peregrine solitons as the initial excitations to the PTNLSE model. In this case we find that under this type of excitation, it always results into two mutually repelling dissipative nonlinear waves due to gradient of the phase distortions [47].

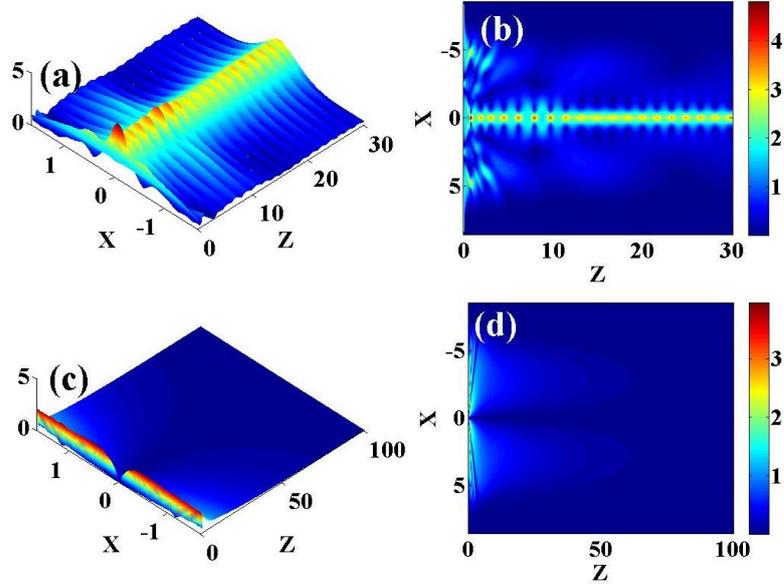

Fig. 1 (Color online): (a) optical fields ($|\psi(x,z)|$) dynamics in the $X - Z$ plane, (b) corresponding density plot of (a). Here $l = 0$, and the interval factor parameter $\varepsilon = 2.5$. Kuznetsov-Ma (KM) soliton train is observed. (c) optical fields ($|\psi(x,z)|$) dynamics in the $X - Z$ plane, (d) corresponding density plot of (c). Here $l = 1$ and the interval factor parameter $\varepsilon = 2.5$. The optical fields of the Peregrine solitons repel each other.

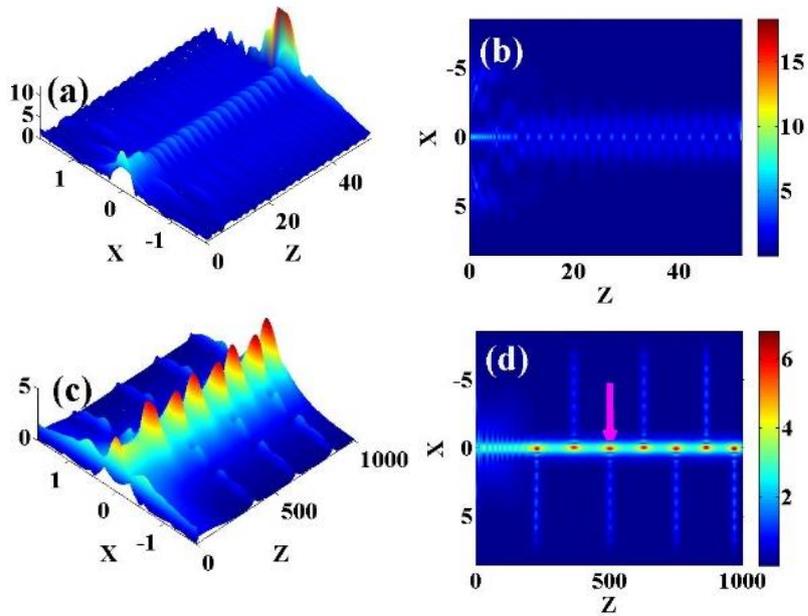

Fig. 2 (Color online): Excitation of a single Peregrine rogue wave: (a) optical fields ($|\psi(x,z)|$) dynamics in the $X - Z$ plane, (b) corresponding density plot of (a). Here $l = 0$ and interval factor parameter $\varepsilon = 0.0$. A string of Peregrine rogue waves excitation: (c) optical fields ($|\psi(x,z)|$) dynamics in the $X - Z$ plane, (d) corresponding density of (c). Here $l = 0$ and the interval factor parameter $\varepsilon = 2.793$. It can be seen from the plots (a),(b) that the KM soliton trains beyond a certain normalized propagation distance to giving rise to single near-ideal Peregrine rogue wave, while a string of Peregrine rogue waves in (c), (d).

Under the set of parameter values chosen, we observe string of highly-localized optical fields dynamics as can be seen in 2(c), (d) over a long normalized propagation distance. The strong spatio-temporal (in $X-Z$ plane) localization characterizes such waves as the near-ideal Peregrine rogue waves. A close inspection into the periodic occurrences of these nonlinear waves further evokes oscillatory pulsations of the power curves in '$z$'. In one of our works [58] we have shown the existence of single Peregrine rogue wave taking a single SFB ansatz as initial condition as opposed to two first order Peregrine solitons in the present work which results in the formation of string of Peregrine rogue waves. The string of highly localized near-ideal Peregrine rogue wave solutions starts to appear when the interval factor parameter takes on a value around $\varepsilon \approx 2.786$. This sort of periodic near-ideal Peregrine rogue wave solution exists in the range of $\varepsilon \in [2.286 - 2.793]$, and for an infinitesimal increase in the value of $\varepsilon$ the solutions enter into a state of blow up. This points towards the fact that rogue waves in general and Peregrine rogue waves in particular are rare and unstable phenomena. The emergence of such waves could be attributed to the nonlinear interaction of two in phase initial Peregrine solitons and modulation instability. Moreover, in ref. [59] it was argued that the interaction between a continuous wave background and a single localized peak perturbation can excite a Peregrine rogue wave. In a similar note, our initial condition in Eq. (5) refers to the same type of initial excitations found to be the precursor of the Peregrine rogue waves but interestingly in PTNLSE system. In Fig. 3 the transverse profile of the optical fields depicts the emergence of the Peregrine rogue waves at various propagation distances under the same parameter values.

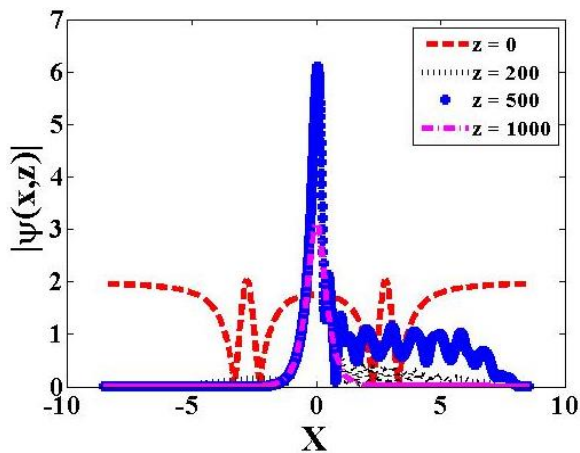

Fig. 3 (Color online): Optical fields ($|\psi(x,z)|$) dynamics of the two first order in phase Peregrine solitons for different values of the normalized propagation distances with the interval factor parameter taken as $\varepsilon = 2.793$. Here we show how the transverse profile changes in the course of evolution. As the value of $z$ increases we see increasing wave localization behavior which at $z = 500$ it reaches its maximum (Peregrine rogue wave) and afterwards becomes unstable. It is the point denoted by the magenta arrow in Fig. 2(d).

*3.2 Effects of the interval factor in the in phase and out of phase conditions*

In Figs. (4) and (5), we have depicted the effects of the interval factor parameter upon the optical fields dynamics. In Fig. (4), for in phase condition, the initial two Peregrine solitons in course of their evolutions attract and then repel each other giving rise to a periodic breathing dynamics. With the decrease of the interval factor parameter in Fig. 4 (d) we observe appearance of a stronger central lobe along with two comparatively weaker side lobes, which could be ascribed to energy transfer from the side lobes to the weaker central lobes as shown in Figs. 4 (b), (c). After forming the fused spot at around $Z \approx 36$, due to strong repulsive force it gives rise to two weaker side lobes propagating with outward transverse velocities and a strong central lobe. The two side lobes gradually disappear beyond $Z \approx 50$ losing energy and only the central lobe persists (see Fig. A2 in appendix A). It is apparent that with the decrease of the transverse shift parameter it takes longer normalized propagation distance ($\Delta z$) for the fusion of the two initial Peregrine solitons to occur. It could be attributed to the nonlocal nonlinear interactions taking place between them due to nonlocal PT-symmetric nonlinearity and its phase modulations. This value of $\Delta z$ varies along $z$. On the other hand, in the out of phase condition, the dissipative oblique propagation dynamics is more pronounced with the decrease of the interval factor parameter.

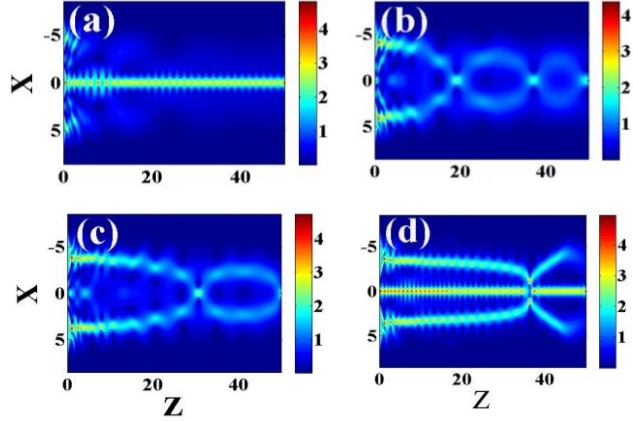

Fig. 4 (Color online): In these density plots we have shown the evolutions of the optical fields dynamics in the $X - Z$ plane in the in phase condition for different values of the interval factor parameter. (a) $\varepsilon = 2.5$, (b) $\varepsilon = 1.5$, (c) $\varepsilon = 1.0$, (d) $\varepsilon = 0.5$. The corresponding transverse profiles showing better initial positions of the input solitons are shown in appendix A.

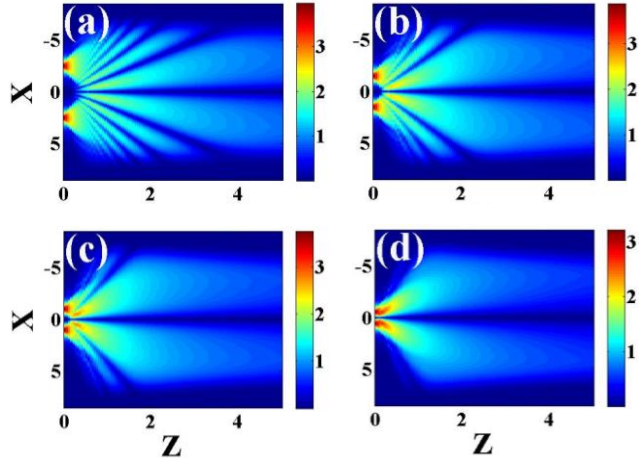

Fig. 5 (Color online): In these density plots we have shown the evolutions of the optical fields ($|\psi(x,z)|$) dynamics in the $X - Z$ plane for out of phase condition for different values of the interval factor parameter. (a) $\varepsilon = 2.5$, (b) $\varepsilon = 1.5$, (c) $\varepsilon = 1.0$, (d) $\varepsilon = 0.5$.

The two initial Peregrine solitons propagate shifting toward the $+X$ and $-X$ directions as is evident from Fig. 5. This could be due to the mutually opposite signs of the induced equivalent wavevector $\Delta k_x$ arising out of the gradient of the phase distortions of the solitons [47].

In Fig. 4 (d) we find the existence of two symmetric sidelobes and one central main lobe. For the same (short) range of the normalized propagation distance the breathing dynamics becomes less pronounced with the increase of the interval factor parameter and the energy transfers to the central main lobe. In the out of phase condition (see Fig. (5)), the Peregrine solitons repel each other over all the values of the interval factor parameter. We also note that as the interval factor parameter is decreased the repulsion between the two initial Peregrine solitons becomes more pronounced due to strong nonlinear interactions. In the in phase condition, the localization of the optical fields has been vividly presented in Fig. 6.

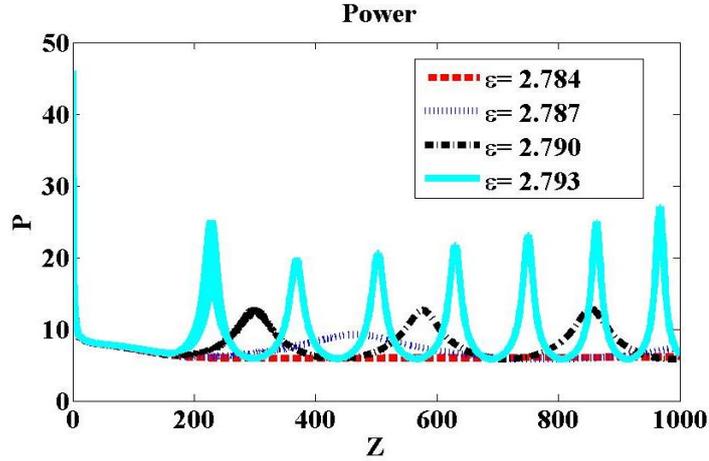

Fig. 6 (Color online): The optical power along the normalized propagation distance for different values of the interval factor parameter for in phase condition to elaborate more on the behaviors of the optical fields dynamics as revealed in Fig. 3. Increasing wave localization behavior can be seen with a small increase $\varepsilon$.

Furthermore, in contrast to the wave dynamics in the standard NLSE, in the PTNLSE model (in phase condition) after fusion of the solitons, the resulting central lobe gives rise to KM soliton trains.

*3.3 Effects of the transverse shift parameter $\varepsilon_{tsp}$: wave instability*

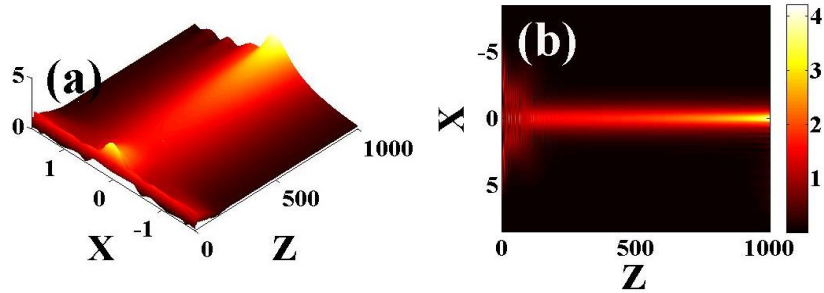

Fig. 7 (Color online): Evolutions of the optical fields ($|\psi(x,z)|$) dynamics in the $X - Z$ plane for in phase condition with a nonzero $\varepsilon_{tsp}$. (a) surface plot, (b) density plot. Values of different parameters chosen are: interval factor parameter $\varepsilon_{loc} = 2$, the transverse shift parameter $\varepsilon_{tsp} = 0.82 \times 10^{-6}$.

In Fig. (7) we have demonstrated the effects of the transverse shift parameter upon the wave dynamics. Under the chosen parameter values we see that the optical fields undergoes instability behavior despite a very small value of $\varepsilon_{tsp}$. It can be attributed as follows: due to the asymmetric field distribution at a given normalized propagation distance if the anti-symmetric imaginary part of the self-induced potential (essentially $V_I(x)$ in $V(x) = V_R(x) + i\, V_I(x)$) goes beyond a certain critical value, the optical fields dynamics can witness the onset of wave instability [32].

## 4. Conclusion and discussions

In this work, we have studied interaction dynamics of two first order Peregrine solitons, in the in phase and out of phase conditions, in the nonlocal continuous nonlinear Schrödinger system with parity-time symmetric Kerr nonlinearity (PTNLSE). Upon numerical computation, in unbroken PT phase we observe the appearance of Kuznetsov-Ma (KM) soliton trains in the in phase condition and repulsive nonlinear waves in the out of phase condition. We also note that the interval factor parameter and the transverse shift parameter do play an important role in the evolution dynamics of the optical fields. Within a specific range of the values of the interval factor parameter $\varepsilon \in$ [2.286 − 2.793], string of highly localized near-ideal Peregrine rogue wave solutions is found to exist and an infinitesimal increase in the value of $\varepsilon$ causes the solutions enter into a state of blow up pointing towards the unstable nature of Peregrine rogue waves. The transverse shift parameter results in wave instability behavior in the optical fields dynamics in the broken PT-phase. It should be noted here that despite the fact that the nonlocal nonlinearity can be found in many physical systems [60, 61], still realizing nonlocal PT symmetric nonlinearity is still elusive, although wave mixing in some proper PT settings has been suggested [32]. Nonetheless, wave localization behavior draws particular interests both fundamentally and technologically, and in that direction, this could enrich the fundamental understanding of the studies pertaining to the emerging field of nonlinear PT optics, particularly to the nonlocal NLSE systems.

**Appendix A**

Here the in Fig. A1 the transverse profiles of the optical fields corresponding to Figs. 4 (a)-(d) have been shown. It should be noted that the changes of the initial locations of the two first order Peregrine solitons are more visible compared to Figs. 4 (a)-(d). In Fig. A2 the optical fields dynamics has been depicted for longer normalized propagation distance corresponding to Fig. 4 (d).

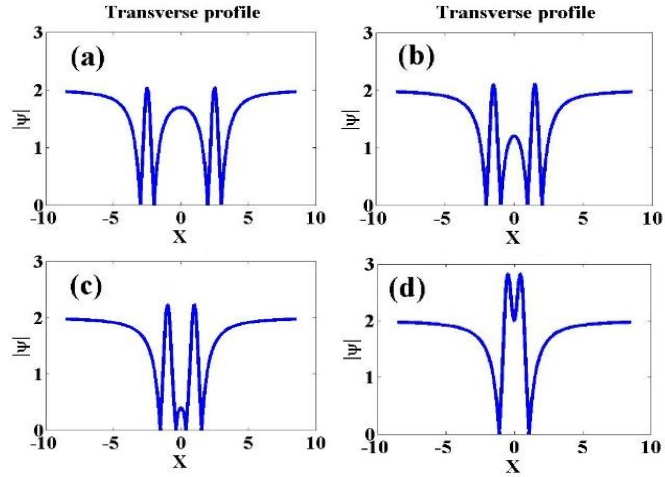

Fig. A1: Transverse field profiles for the in phase condition corresponding to Fig. 4 in the paper. (a) $\varepsilon = 2.5$, (b) $\varepsilon = 1.5$, (c) $\varepsilon = 1.0$, (d) $\varepsilon = 0.5$.

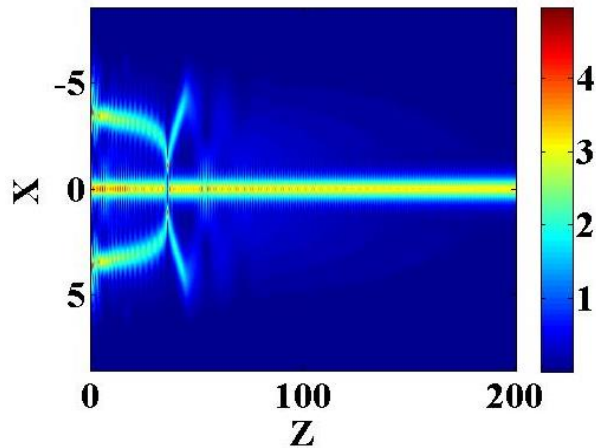

Fig. A2: Optical field dynamics in the in phase condition corresponding to Fig. 4 (d) for $\varepsilon = 0.5$ for longer propagation distance.

**Acknowledgments**

S. K. G. would like to thank the anonymous reviewers for their constructive suggestions which helped in improving this work. This research did not receive any specific grant from funding agencies in the public, commercial, or not-for-profit sectors.

**References**

[1] C.M. Bender, S. Boettcher, Real spectra in Non-Hermitian Hamiltonian Having PT-Symmetry, Phys. Rev. Lett. 80 (1998) 5243-5246.
[2] K.G. Makris, R. El-Ganainy, D. N. Christodoulides, Z. H. Musslimani, Beam Dynamics in PT-symmetric Optical Lattices, Phys. Rev. Lett. 100 (2008) 103904-07.
[3] A. Guo et al., Observation of PT-symmetry Breaking in Complex Optical Potentials, Phys. Rev. Lett. 103 (2009) 093902-05.
[4] C. E. Rüter, K. G. Makris, R. El-Ganainy, D. N. Christodoulides, M. Segev, D. Kip, Observation of parity-time symmetry in optics, Nat. Phys. 6 (2010) 192-195.
[5] A. Regensburger, C. Bersch, M.–A. Miri, G. Onishchukov, D. N. Christodoulides, U. Peschel, Parity-time synthetic photonic lattices. Nature 488 (2012) 167-171.
[6] R. El-Ganainy, K.G. Makris, D.N. Christidoulides, Z.H. Musslimani, Theory of coupled optical PT-symmetric structures, Opt. Lett. 32 (2007) 2632-34.
[7] A. Regensburger, M.–A. Miri, C. Bersch, J. Nager, G. Onishchukov, D.N. Christodoulides, U. Peschel, Observation of Defect States in PT-symmetric Optical Lattices, Phys. Rev. Lett. 110 (2013) 223902-06.
[8] M.-A. Miri, A. Regensburger, U. Peschel, D.N. Christodoulides, Optical Mess Lattices with PT-symmetry, Phys. Rev. A 86 (2012) 023807-18.
[9] M.C. Zheng, D.N. Christodoulides, R. Fleischmann, T. Kottos, PT optical lattices and universality in beam dynamics, Phys. Rev. A 82 (2010) 010103-06.
[10] S. Longhi, Bloch Oscillations in Complex Crystals with PT symmetry, Phys. Rev. Lett. 103 (2009) 123601-04.
[11] Y.D. Chong, L. Ge, A.D. Stone, PT-symmetry Breaking and Laser–Absorber Modes in Optical Scattering Systems, Phys. Rev. Lett. 106 (2011) 093902-05.
[12] A. Szameit, M.C. Rechtsman, O. Bahat-Treidei, M. Segev, PT-symmetry in honeycomb photonic lattices, Phys. Rev. A 84 (2011) 021806-10.
[13] E.–M. Graefe, H.F. Jones, PT-symmetric sinusoidal optical lattices at the symmetry breaking threshold, Phys. Rev. A 84 (2010) 013818-25.
[14] S. Longhi, PT-symmetric laser absorber, Phys. Rev. A 82 (2010) 031801-04.
[15] Y.N. Joglekar, J.L. Barnett, Origin of maximal symmetry breaking in even PT-symmetric lattices, Phys. Rev. A 84 (2011) 024103-05.
[16] M.–A. Miri, P. LikamWa, D.N.Christodoulides, Large area single-mode parity-time-symmetric laser amplifiers, Opt. Lett. 37 (2012) 764-6.
[17] M. Liertzer, L. Ge, A. Cerjan, A.D. Stone, H.E. Turcei, S. Rotter, Pump-induced Exceptional Points in Lasers, Phys. Rev. Lett. 108 (2012) 173901-05.
[18] B. Midya, B. Roy, R. Roychoudhuri, A note on the PT invariant potential $(x) = 4cos^2(x) + 4iV_0 \sin(2x)$, Phys. Lett. A 374 (2010) 2605-07.
[19] M.–A. Miri, M. Heinrich, D.N. Christodoulides, Supersymmetry-generated complex optical potentials with real spectra, Phys. Rev. A 87 (2013) 043819-23.
[20] S.K. Gupta, A.K. Sarma, Solitary waves in parity-time (PT)-symmetric Bragg grating structure and the existence of optical rogue waves, Europhys. Lett. 105 (2014) 44001-06.
[21] H. Ramezani, T. Kottos, R. El-Ganainy, D.N. Christodoulides, Unidirectional nonlinear PT-symmetric optical structures, Phys. Rev. A 82 (2010) 043803-08.
[22] K. Li, P.G. Kevrekidis, PT-symmetric oligomers: Analytical solutions, linear stability, and nonlinear dynamics, Phys. Rev. E 83 (2011) 066608-14.
[23] M. Duanmu, K. Li, P.G. Kevrekidis, N. Whitaker, Linear and nonlinear parity-time-symmetric oligomers: a dynamical systems analysis. Philos. Trans. R. Soc. A 371 (2013) 20120171-89.
[24] N.V. Alexeeva, I.V. Barashenkov, A.A. Sukhorukov, Y.S. Kivshar, Optical solitons in PT-symmetric nonlinear couplers with gain and loss, Phys. Rev. A 85 (2012) 063837-49.
[25] M.–A. Miri, A.B. Aceves, T. Kottos, V. Kovanis, D.N. Christodoulides, Bragg solitons in PT-symmetric periodic potentials, Phys. Rev. A 86 (2012) 033801-05.
[26] S.V. Suchkov, B.A. Malomed, S.V. Dmitriev, Y.S. Kivshar, Solitons in a chain of parity-time invariant dimers, Phys. Rev. E 84 (2011) 046609-16.
[27] A.A. Sukhorukov, Z. Xu, Y.S. Kivshar, Nonlinear suppression of time reversal in PT-symmetric optical couplers, Phys. Rev. A 82 (2010) 043818-22.
[28] Z.H. Musslimani, K.G. Makris, R. El-Ganainy, D.N. Christodoulides, Optical solitons in PT periodic potentials, Phys. Rev. Lett. 100 (2008) 030402-05.
[29] R. Driben, B.A. Malomed, Stability of solitons in parity-time-symmetric couplers, Opt. Lett. 36 (2011) 4323-25.
[30] YV Bludov, VV Konotop, BA Malomed, Stable dark solitons in PT-symmetric dual-core waveguides, Phys. Rev. A 87 (2013) 013816-22.
[31] M.J. Ablowitz, Z.H. Musslimani, Integrable Nonlocal Nonlinear Schrödinger Equation, Phys. Rev. Lett. 110 (2013) 064105-09.
[32] A.K. Sarma, M.–A. Miri, Z.H. Musslimani, D.N. Christodoulides, Continuous and discrete Schrödinger systems with parity-time-symmetric nonlinearity, Phys. Rev. E 89 (2014) 052918-24.

[33] M. Li, T. Xu, Dark and anti-dark soliton interaction in the nonlinear nonlocal Schrödinger equation with self-induced parity-time-symmetric potential, Phys. Rev. E 91 (2015) 033202-09.

[34] A Khare, A Saxena, Periodic and hyperbolic soliton solutions of a number of nonlocal nonlinear equations, J. Math. Phys. 56 (2015) 084101.

[35] X.-Y. Wen, Z. Yan, Y. Yang, Dynamics of higher order rational solitons for the nonlocal nonlinear Schrödinger equation with self-induced parity-time-symmetric potential, Chaos 26 (2016) 063123.

[36] M. Li, T. Xu, D. Meng, Rational Solitons in the Parity-Time-Symmetric Nonlocal Nonlinear Schrödinger Model, J. Phys. Soc. Jpn. 85 (2016) 124001.

[37] T. Xu, H. Li, H. Zhang, M. Li, S. Lan, Darboux transformation and analytic solutions of the discrete PT-symmetric nonlocal nonlinear Schrödinger equation, 63 (2017) 88-94.

[38] D Sinha, PK Ghosh, Symmetries and exact solutions of a class of nonlocal nonlinear Schrödinger equations with self-induced parity-time-symmetric potential. Phys. Rev. E 91 (2015) 042908-22.

[39] P.S. Vinayagam, R. Radha, U.A. Khawaja, L. Ling, Collisional dynamics of solitons in the coupled PT symmetric nonlocal nonlinear Schrödinger equations, Commun. Nonlin. Sci. Numer. Simulat. 52 (2017) 1-10.

[40] B. Kibler, J. Fatome, C. Finot, G. Millot, F. Dias, G. Genty, N. Akhmediev, J.M. Dudley, The Peregrine soliton in nonlinear fiber optics, Nat. Phys. 6 (2010) 790-95.

[41] C. Dai, Y. Wang, X. Zhang, Controllable Akhmediev breather and Kuznetsov-Ma soliton trains in PT-symmetric coupled waveguides, Opt. Exp. 22 (2014) 29862-67.

[42] J.M. Dudley, G. Genty, F. Dias, B. Kibler, N. Akhmediev, Modulation instability, Akhmediev breather and continuous wave supercontinuum generation, Opt. Exp. 17 (2009) 21497-21508.

[43] B. Kibler, J. Fatome, C. Finot, G. Millot, G. Genty, B. Wetzel, N. Akhmediev, F. Dias, J.M. Dudley, Observation of Kuznetsov-Ma soliton dynamics in optical fibre, Sci. Rep. 2 (2012) 463.

[44] D.H. Peregrine, Water waves, nonlinear Schrödinger equation and their solutions, J. Aust .Math. Soc. Ser. B 25 (1983) 16-43.

[45] H. Bailung, S.K. Sharma, Y. Nakamura, Observation of Peregrine Solitons in a Multi-component Plasma with Negative Ions, Phys. Rev. Lett. 107 (2011) 255005-08.

[46] K. Hammani, B. Kibler, C. Finot, P. Morin, J. Fatome, J.M. Dudley, G. Millot, Peregrine solitons generation and break-up in standard telecommunications fiber, Opt. Lett. 36 (2011) 112-114.

[47] W. Zhen-Kun, Z. Yun-Zhe, H. Yi, W. Feng, Z. Yi-Qi, Z. Yan-Peng, The Interactions of Peregrine Solitons, Chin. Phys. Lett. 31 (2014) 090502.

[48] G. Yang, Y. Wang, Z. Qin, B.A. Malomed, D. Mihalache, L. Li, Breatherlike solitons extracted from the Peregrine rogue waves, Phys. Rev. E 90 (2014) 062909-15.

[49] S. Chen, L. Song, Peregrine solitons and algebraic soliton pair in Kerr media considering space-time correction, Phys. Lett. A 378 (2014) 1228-32.

[50] V.I. Shrira, V.V. Geogjaev, What makes the Peregrine solitons so special as a prototype of freak waves? J. Eng. Math. (2010) 67 11-22.

[51] A.I. Dyachenko, V.E. Zakharov, Modulation instability of Stokes waves→Freak waves, JETP Lett. 81 (2005) 255-259.

[52] A. Chabchoub, N.P. Hoffmann, N. Akhmediev, Rogue Wave Observation in a Water Wave Tank. Phys. Rev. Lett. 106 (2011) 204502-05.

[53] Y.V. Bludov, V.V. Konotop, N. Akhmediev, Matter rogue waves, Phys. Rev. A 80 (2009) 033610-15.

[54] G. Oppo, A.M. Yao, D. Cuozzo, Self-organization, pattern formation, cavity solitons and rogue waves in singly resonant optical parametric oscillators. Phys. Rev. A 88 (2013) 043813-23.

[55] J. Zamora-Munt, B. Garbin, S. Barland, M.Giudici, J.R.R. Leite, C. Masoller, J.R. Tredicce, Rogue waves in optically injected lasers: origin, predictability and suppression, Phys. Rev. A 87 (2013) 035802-07.

[56] A Zaviyalov, O Egorov, R Iliew, F Lederer, Rogue waves in mode-locked fiber lasers, Phys. Rev. A 85 (2012) 013828-14.

[57] G. P. Agrawal, Nonlinear Fiber Optics, 5th ed., Academic Press, San Diego, 2013.

[58] S.K. Gupta, A.K. Sarma, Peregrine rogue wave dynamics in the continuous nonlinear Schrödinger system with parity-time symmetric Kerr nonlinearity, Commun. Nonlin. Sci. Numer. Simulat. 36 (2016) 141-147.

[59] G. Yang, L. Li, S. Jia, Peregrine rogue waves induced by the interaction between a continuous wave and a soliton, Phys. Rev. E 85 (2012) 046608-16.

[60] S. P. Cockburn, H. E. Nistazakis, T. P.Horikis, P. G.Kevrekidis, N. P.Proukakis, D. J.Frantzeskakis, Fluctuating and dissipative dynamics of dark solitons in quasicondensates, Phys. Rev. Lett. 84 (2011) 043640.

[61] F. Maucher, E. Siminos, W. Krolikowski, S. Skupin, Quasiperiodic oscillations and homoclinic orbits in the nonlinear nonlocal Schrodinger equation, New J. Phys. 15 (2013) 083055.